\documentclass{JHEP3}

\usepackage{Pformel,Pmeta,Ppicture}
\usepackage{amsmath,amsfonts,amsthm,times}

\preprint{LPTA/09-100}

\title{Minisuperspace limit of the $AdS_3$ WZNW model}

\author{Sylvain Ribault
\\ 
 Laboratoire de Physique Th\'eorique et Astroparticules, UMR5207 CNRS-UM2,
 \\
 Universit\'e Montpellier II, Place E. Bataillon, CC 070
 \\
 34095 Montpellier Cedex 05, France 
 \\
 {\footnotesize \tt sylvain.ribault@um2.fr }
}

\abstract{We derive the three-point function of the $AdS_3$ WZNW model in the minisuperspace limit by Wick rotation from the $\Hp$ model. The result is expressed in terms of Clebsch-Gordan coefficients of the Lie algebra $s\ell(2,\R)$. We also introduce a covariant basis of functions on $AdS_3$, which can be interpreted as bulk-boundary propagators. 
}

\makeatletter\let\default@color\current@color\makeatother 

\begin{document}

\zeq\section{Introduction}

The $AdS_3$ Wess--Zumino--Novikov--Witten model is interesting in particular due to its string theory applications. A conjecture for the spectrum of this model was proposed by Maldacena and Ooguri \cite{mo00a}, but the full solution of the model is still missing. In the sense of the conformal bootstrap, a full solution means the computation of the three-point functions of primary fields on the sphere, and the proof of crossing symmetry of the four-point functions. (Equivalently, the computation of operator product expansions of primary fields, and the proof of their associativity.) 

The conjectured spectrum of the $AdS_3$ WZNW model is fairly complicated, as it contains both discrete and continuous series of representations of the symmetry algebra, and their images under the so-called spectral flow automorphism. 
On the other hand, as a geometrical space, $AdS_3$ is related by Wick rotation to the Euclidean space $\Hp$, and the $AdS_3$ WZNW model is often assumed to be related to the $\Hp$ WZNW model. The spectrum of the latter model is much simpler, as it contains only a continuous series of representations, and the $\Hp$ model has been fully solved \cite{tes99,tes01b}. An additional difficulty of the $AdS_3$ WZNW model is that the group $AdS_3$, which is the universal cover of $\SLR$, has no realization as a group of finite-dimensional matrices. It follows that writing a simple basis of functions on $AdS_3$ is more difficult than in the cases of $\SLR$ or $\Hp$. Similarly, it is in general more complicated to write functions on the Anti-de Sitter space $AdS_d$ than on its Euclidean version $H_d^+$. Some works like \cite{wit98} which are purportedly about $AdS_d$ actually deal with $H_d^+$, thereby avoiding this difficulty (and other ones). 

The presence of discrete representations and the lack of an obvious basis of functions on $AdS_3$ are two difficulties of the $AdS_3$ WZNW model which also affect its minisuperspace limit (also known as the zero-mode approximation), where the model reduces to the study of functions on the $AdS_3$ space. It is therefore interesting to solve the model in this limit. We will do this by starting from the well-understood minisuperspace $\Hp$ model \cite{tes97b} and using Wick rotation. The main object we wish to compute is the minisuperspace analog of the operator product expansion, namely the product of functions on $AdS_3$. (Equivalently, the minisuperspace three-point function.)

We will start with a study of certain bases of functions on $AdS_3$, $\SLR$ and $\Hp$ (Section 2). In particular we will construct functions on $AdS_3$ which transform covariantly under the symmetries, and which can be interpreted as bulk-boundary propagators. Then we will study the Clebsch-Gordan coefficients of the Lie algebra $s\ell(2,\R)$ (Section 3). Due to the symmetries of the $AdS_3$ WZNW model in the minisuperspace limit, the products of functions on $AdS_3$ can be expressed in terms of these coefficients. We will check this after obtaining these products of functions by Wick rotation from $\Hp$ (Section 4). In conclusion we will comment on the Wick rotation and on the problem of solving the $AdS_3$ WZNW model (Section 5).

\zeq\section{$\Hp$, $\SLR$, $AdS_3$ and functions thereon}

In this section we will review the geometries of the spaces $\Hp$, $\SLR$ and $AdS_3$, and introduce bases of functions on these spaces. The sense in which such functions form bases of certain functional spaces will be  explained in section \ref{secbas}.
While $\Hp$ and $\SLR$ can be viewed as spaces of two-dimensional matrices, $AdS_3$ cannot, and this will make the descriptions of functions on $AdS_3$ more complicated. 

\subsection{Geometry and symmetry groups}

Let us start with the group $\SLR$ of real, size two matrices of determinant one. This group is not simply connected, since the subgroup of the matrices 
\bea
g_\tau\equiv\bsm \cos \tau & \sin \tau \\ -\sin \tau & \cos \tau \esm
\label{gtau}
\eea 
is a non-contractible loop. Therefore, there exists a universal covering group, sometimes called $\widetilde{SL}(2,\R)$, which we will call $AdS_3$. If $\SLR$ elements are parametrized using three real coordinates $(\rho,\theta,\tau)$ as
\bea
g= \left(\begin{array}{ccc} \cosh \rho \cos \tau + \sinh \rho \cos \theta & & \sinh \rho \sin \theta +\cosh\rho \sin\tau \\ \sinh \rho \sin \theta - \cosh \rho \sin \tau & & \cosh \rho \cos \tau -\sinh \rho \cos \theta
\end{array}\right)\ ,
\label{amat}
\eea
where $\theta$ and $\tau$ are $2\pi$-periodic, then $AdS_3$ is obtained by decompactifying $\tau$. 
Elements of $AdS_3$ can alternatively be parametrized as doublets $G=(g,I)$ where $g\in \SLR$ and $I$ is the integer part of $\frac{\tau}{2\pi}$. Writing $\frac{\tau}{2\pi}=I+F$, the group multiplication of $AdS_3$ can be written as 
\bea
(g,I)(g',I')=(gg',I+I'+F(g)+F(g')-F(gg'))\ .
\eea
The $U(1)$ subgroup of the matrices $g_\tau\in\SLR$ decompactifies into an $\R$ subgroup of elements $G_\tau\in AdS_3$, which we parametrize as
\bea
G_\tau = \exp \tau \bsm 0 & 1 \\ -1 & 0 \esm\ .
\label{gltau}
\eea
The group structures of $AdS_3$ and $\SLR$ lead to left and right actions by group multiplication, in the  $AdS_3$ case $(G_L,G_R)\cdot G=G_LGG_R$, which will be symmetries of the models under consideration. More precisely, the geometrical symmetry group of $\SLR$ is $\frac{\SLR\times \SLR}{\Z_2}$, where we must divide by the center $\Z_2=\{\id,-\id\}$ of $\SLR$ as its left and right actions are identical. The geometrical symmetry group of $AdS_3$ is $\frac{AdS_3\times AdS_3}{\Z}$, where the center of $AdS_3$ is freely generated by $(-\id,0)=(\rho=0,\theta=0,\tau=\pi)$
and therefore isomorphic to $\Z$. These geometrical symmetry groups are not simply connected; their first fundamental groups are $\Z^2$ in the case of $\SLR$ and $\Z$ in the case of $AdS_3$. This is the origin of the spectral flow symmetries of the corresponding WZNW models. (See for instance \cite{mo00a}.)

Then $\Hp$ is the space of hermitian, size two matrices of determinant one, which can be parametrized using three real coordinates $(\rho,\theta,\tau)$ as
\bea
h=\left(\begin{array}{cc} e^\tau \cosh \rho & e^{i\theta}\sinh \rho \\ e^{-i\theta}\sinh \rho & e^{-\tau}\cosh \rho \end{array}\right) \ .
\label{hmat}
\eea
We have chosen identical names $(\rho,\theta,\tau)$ for the coordinates on $\Hp$ and $AdS_3$, thereby defining a bijection between these two spaces. This bijection gives rise to a map $\Phi(\rho,\theta,\tau)\rar \Phi(\rho,\theta,i\tau)$ from the analytic functions on $\Hp$ to the analytic functions on $AdS_3$, which is called the Wick rotation. Our bijection however does not relate the matrix forms of $\Hp$ and $\SLR$ which we have given. Notice that $\Hp$
is not a group, rather a group coset, namely $\SLC/SU(2)$. The geometrical symmetry group of $\Hp$ is $\frac{\SLC}{\Z_2}$, whose elements $k$ act on $h\in \Hp$ by $k\cdot h = khk^\dagger$.

\subsection{Functions: $t$-bases \label{ssect}}

In both cases $\SLR$ and $\Hp$, the existence of a matrix realization allows us to write functions which transform very simply under the symmetries. In the case of $\Hp$, we can indeed introduce the following ``$x$-basis'' of functions $\Phi^j_x(h)$, parametrized by their spin $j\in\C$ and isospin $x\in \C$:
\bea
\Phi^j_x(h)\equiv \frac{2j+1}{\pi} \left|(\bx,1) h \bsm x \\ 1 \esm \right|^{2j}\ \  \Rightarrow \ \ \Phi^j_x(khk^\dagger) = |cx+d|^{4j} \Phi^j_{\frac{ax+b}{cx+d}}(h)\ ,
\eea
where we denote $k^\dagger = \bsm a & b \\ c & d \esm$. Similarly, we can introduce the following ``$t$-basis''
of functions $\Phi^{j,\eta}_{t_L,t_R}(g)$ on $\SLR$, with $(t_L,t_R)\in\R^2$:
\begin{multline}
\Phi^{j,\eta}_{t_L,t_R}(g) \equiv \frac{2j+1}{\pi} \left|(1,-t_L)g \bsm t_R \\ 1 \esm \right|^{2j} \sign^{2\eta} (1,-t_L)g \bsm t_R \\ 1 \esm 
\ \  \Rightarrow \ \ 
\Phi^{j,\eta}_{t_L,t_R}(g_L^{-1} g g_R) 
\\
= \left|(c_Rt_R+d_R)(c_Lt_L+d_L)\right|^{2j} \sign^{2\eta}(c_Rt_R+d_R)(c_Lt_L+d_L)\ \Phi^{j,\eta}_{\frac{a_Lt_L+b_L}{c_Lt_L+d_L},\frac{a_Rt_R+b_R}{c_Rt_R+d_R}}(g)\ , 
\label{pjeg}
\end{multline}
where we denote $g_L=\bsm a_L & b_L \\ c_L & d_L \esm $ and $g_R=\bsm a_R & b_R \\ c_R & d_R\esm$. The parity $\eta\in\{0,\frac12\}$ is the same for both actions of $\SLR$ on itself by multiplications from the left and from the right (see the factor $\sign^{2\eta}(c_Rt_R+d_R)(c_Lt_L+d_L)$), because the parity characterizes the action of the central subgroup $\Z_2$. 

In the case of $AdS_3$, writing a similar ''$t$-basis`` of functions is more complicated.
We define $t$-basis functions $\Phi^{j,\al}_{t_L,t_R}$ on $AdS_3$ by the assumption that they transform covariantly under the left and right actions of $AdS_3$ on itself, in a way which generalizes
the transformation property of the $t$-basis functions $\Phi^{j,\eta}_{t_L,t_R}$ on $\SLR$ eq. (\ref{pjeg}). (The $AdS_3$ parameter $\al\in[0,1)$ generalizes the $\SLR$ parameter $\eta\in\{0,\frac12\}$.)
The appropriate generalization of the transformation property has been written in \cite{hr06} (Section 4.1); it involves a function $N(G|t)$ on $AdS_3\times \R$ such that $N(G'G|t)=N(G'|Gt)+N(G|t)$ and $\forall n\in\Z,\ N((\id,I)|t)=I$, where if $G=(g,I)=(\bsm a & b \\ c & d\esm,I)$ then $Gt\equiv gt= \frac{at+b}{ct+d}$. For instance, $N(G|t)$ can be taken as the number of times $G't$ crosses infinity as $G'$ moves from $(\id,0)$ to $G$, in which case $N(G|t)\in \Z$, and $[G]\equiv N(G|t)-\frac12\sign(t+\frac{d}{c})-\frac12$ is a $t$-independent integer. Then, the axiom for $\Phi^{j,\al}_{t_L,t_R}$ is
\bea
\Phi^{j,\al}_{t_L,t_R}(G_L^{-1} G G_R) 
= \left|(c_Rt_R+d_R)(c_Lt_L+d_L)\right|^{2j} e^{2\pi i \al (N(G_L|t_L)-N(G_R|t_R))} \Phi^{j,\al}_{G_Lt_L,G_Rt_R}(G)\ .
\label{pnnp}
\eea
This axiom is obeyed by 
\bea
\Phi^{j,\al}_{t_L,t_R}(G) =\frac{2j+1}{\pi} e^{2\pi i \al n(G|t_L,t_R)} \left|(1,-t_L)g \bsm t_R \\ 1 \esm \right|^{2j} \ ,
\label{pjatt}
\eea
provided the function $n(G|t_L,t_R)$ satisfies 
\bea
n(G_L^{-1}GG_R|t_L,t_R)-n(G|G_Lt_L,G_Rt_R) = N(G_L|t_L)-N(G_R|t_R)\ .
\eea
This implies that the function $n(\id|t_L,t_R)$ should satisfy
\bea
n(\id|t_L,Gt_R)-n(\id|G^{-1}t_L,t_R)=N(G^{-1}|t_L)+N(G|t_R)\ ,
\eea
which, using $Gt=gt=\frac{at+b}{ct+d}$ and the properties of $N(G|t)$, amounts to
\bea
n(\id|t_L,\tfrac{at_R+b}{ct_R+d})-n(\id|\tfrac{dt_L-b}{-ct_L+a},t_R)
=\tfrac12 \sign(t_L-\tfrac{a}{c}) + \tfrac12 \sign(t_R+\tfrac{d}{c})\ .
\eea
A solution is found to be 
\bea
n(\id|t_L,t_R)=\tfrac12\sign(t_L-t_R)\ ,
\eea
then $n(G|t_L,t_R)-\tfrac12\sign(t_L-t_R)$ is the number of times $gt_L$ crosses $t_R$ when $g$ runs from $\id$ to $G$. Let us now study the behaviour of $n(G|t_L,t_R)$ as a function of $t_L,t_R$ for a generic choice of $G$. Notice that $\tilde{n}(t_L,t_R)\equiv n(G|t_L,t_R)+[G]+\frac12 = \tfrac12\left[\sign\left((t_L-\tfrac{a}{c})(t_R+\tfrac{d}{c})+\tfrac{1}{c^2}\right)-1\right]\sign(t_R+\tfrac{d}{c})$ takes values $0,\pm 1$, and jumps between these values occur on the hyperbola with equation $(1,-t_L) g \bsm t_R \\ 1 \esm =0$. These values and these jumps are shown on the following plot:
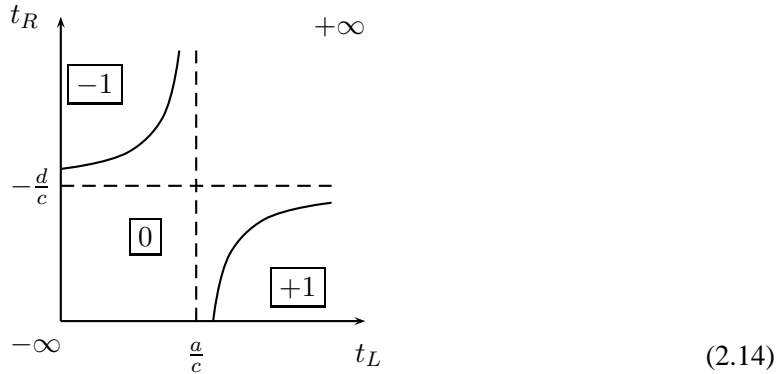
\bea
\psset{unit=.45cm}
\pspicture[](-5,-5.3)(5,5.3)  
\psline{->}(-4,-4)(5,-4)
\psline{->}(-4,-4)(-4,5)
\psline[linestyle=dashed](-4,0)(4,0)
\psline[linestyle=dashed](0,-4)(0,4)
\rput[l]{0}(-5.5,5){$t_R$}
\rput[r]{0}(5.5,-5){$t_L$}
\rput[bl]{0}(-5.5,-5){$-\infty$}
\rput[tr]{0}(5,5){$+\infty$}
\rput[c]{0}(-3,3){$\boxed{-1}$}
\rput[c]{0}(3,-3){$\boxed{+1}$}
\rput*[c]{0}(-1.5,-1.5){$\boxed{0}$}
\rput[l]{0}(-5.5,0){$-\tfrac{d}{c}$}
\rput[b]{0}(0,-5.5){$\tfrac{a}{c}$}
\pscurve(-4,.5)(-2,1)(-1,2)(-.5,4)
\pscurve(4,-.5)(2,-1)(1,-2)(.5,-4)
\endpspicture
\label{tpic}
\eea
We now propose that certain linear combinations of the functions $\Phi^{j,\al}_{t_L,t_R}$ can be interpreted as bulk-boundary propagators. These combinations are
\bea
\Phi^j_{(t_L,t_R,N)}(G)\equiv\int_0^1 d\al\ e^{-2i\pi\al N}\Phi^{j,\al}_{t_L,t_R}(G) =\delta_{N,n(G|t_L,t_R)}\left| (1,-t_L) g \bsm t_R \\ 1 \esm\right|^{2j} \ .
\eea
We interpret $(t_L,t_R,N)\in \R\times \R \times \Z$ as coordinates on the boundary of $AdS_3$. The action of the symmetry group $AdS_3\times AdS_3$ on the boundary is then given by
\bea
(G_L,G_R)\cdot (t_L,t_R,N) = (g_Lt_L,g_Rt_R,N-N(G_L|t_L)+N(G_R|t_R))\ ,
\eea
and the behaviour of $\Phi^{j,\al}_{t_L,t_R}$ under the action of $AdS_3\times AdS_3$ (\ref{pnnp}) implies the following behaviour of $\Phi^j_{(t_L,t_R,N)}$:
\bea
\Phi^j_{(t_L,t_R,N)}(G_L^{-1}GG_R)=\left|(c_Rt_R+d_R)(c_Lt_L+d_L)\right|^{2j} \Phi^j_{(G_L,G_R)\cdot (t_L,t_R,N)}(G)\ . 
\eea

\subsection{Functions: $m$-bases}

The $t$-bases of functions behave simply under symmetry transformations, but $t$-bases in $\Hp$ and $AdS_3$ are not related by the Wick rotation. This is because the matrix realizations (\ref{amat}) and (\ref{hmat}) on which the $t$-bases are built are themselves not related by the Wick rotation. We will therefore introduce
the more complicated ``$m$-bases'' of functions, which are better suited to the Wick rotation. In the case of $\Hp$, the $m$-basis functions $\Phi^j_{m,\bar{m}}(h)$ are defined as 
\bea
\Phi^j_{m,\bar{m}}(h) \equiv \int d^2x\ x^{-j-1+m}\bar{x}^{-j-1+\bar{m}} \Phi^j_x(h) \ \ \ {\rm with} \ \ \ m-\bar{m}\in \Z\ .
\label{phph}
\eea
The numbers $m,\bar{m}$ can be written in terms of an integer $n\in \Z$ and a momentum $p$, which is imaginary in the $\Hp$ model:
\bea
m=\frac12(n+p) \scs \bar{m}=\frac12(-n+p) \ .
\eea
The explicit expression for $\Phi^j_{m,\bar{m}}(h)$ is found to be 
\begin{multline}
\Phi^j_{m,\bar{m}}(h)= -4 \frac{\G(-j+\frac{|n|+p}{2})\G(-j+\frac{|n|-p}{2})}{\G(|n|+1)\G(-2j-1)} e^{-p\tau+in\theta} \sinh^{|n|}\rho\ \cosh^p\rho\ 
\\
\times
F(-j+\tfrac{|n|+p}{2},j+1+\tfrac{|n|+p}{2},|n|+1,-\sinh^2\rho)\ .
\label{pjh}
\end{multline}
Notice that this obeys the so-called reflection property
\bea
\Phi^j_{m,\bar{m}} = R^j_{m,\bar{m}} \Phi^{-j-1}_{m,\bar{m}}\scs R^j_{m,\bar{m}} &=&\frac{\G(2j+1)}{\G(-2j-1)} \frac{\G(-j+m)\G(-j-\bar{m})}{\G(j+1+m) \G(j+1-\bar{m})}
\label{rj}
\\
&=& \frac{\G(2j+1)}{\G(-2j-1)} \frac{\G(-j+\frac{|n|+p}{2})\G(-j+\frac{|n|-p}{2})}{\G(j+1+\frac{|n|+p}{2}) \G(j+1+\frac{|n|-p}{2})}\ ,
\eea
where $R^j_{m,\bar{m}}=R^j_{\bar{m},m}$ due to $n=m-\bar{m}\in\Z$.

We will use the functions on $AdS_3$ obtained from the above functions $\Phi^j_{m,\bar{m}}(h)$ by the Wick rotation $\tau \rar i\tau$. In order for the resulting functions to be delta-function normalizable, we now need to assume the momentum $p$ to be real, instead of imaginary in the $\Hp$ case. We do not introduce a new notation for the resulting functions on $AdS_3$, but still call them $\Phi^j_{m,\bar{m}}(G)$ or $\Phi^j_{m,\bar{m}}$. 

In contrast to $t$-basis functions, $m$-basis functions on $AdS_3$ do not transform simply under the action of the $AdS_3\times AdS_3$ symmetry group. However, they do transform simply under the action of the $\R\times \R$ subgroup made of pairs $(G_{\tau_L},G_{\tau_R})$, where $G_\tau$ was defined by eq. (\ref{gltau}):
\bea
\Phi^{j,\al}_{m,\bar{m}}(G_{\tau_L} G G_{\tau_R}) = e^{-2i(m\tau_L+\bar{m}\tau_R)} \Phi^{j,\al}_{m,\bar{m}}(G)\ .
\label{ptltr}
\eea
Notice that the identity $G_\pi G G_{-\pi} = G$ implies $m-\bar{m}\in \Z$. (In the particular case of $\SLR$, we have the additional identity $g_{2\pi}=\id$, which implies
$m,\bar{m}\in \tfrac12 \Z$.) Introducing 
\bea
\al \in [0,1) \qquad {\rm such\ that} \qquad m,\bar{m}\in \al+\Z\ ,
\eea
this parameter $\al$ is identical to the parameter $\al$ of the $t$-basis functions $\Phi^{j,\al}_{t_L,t_R}$ (\ref{pjatt}). We will look for a relation of the type
\bea
\Phi^j_{m,\bar{m}}=c^{j,\al} \int_\R dt_L\ (1+t_L^2)^{-j-1}e^{i\pi m}\left(\tfrac{1-it_L}{1+it_L}\right)^m \int_\R dt_R\ (1+t_R^2)^{-j-1}\left(\tfrac{1+it_R}{1-it_R}\right)^{\bar{m}} \Phi^{j,\al}_{t_L,t_R}\ ,
\label{pmpt}
\eea
where $c^{j,\al}$ is a normalization factor. We can check that the right-hand side of this relation obeys the transformation property (\ref{ptltr}), thanks to the behaviour eq. (\ref{pnnp}) of $\Phi^{j,\al}_{t_L,t_R}$. To see this it is useful to notice that the integrand in eq. (\ref{pmpt}) is continuous through $t_L=\infty$ and $t_R=\infty$, as can be deduced from the behaviour of the phase factor $e^{2i\pi \al n(G|t_L,t_R)}$ of $\Phi^{j,\al}_{t_L,t_R}$, which is depicted in the diagram (\ref{tpic}). This makes it possible to perform translations of the variables $\varphi_L,\varphi_R$ such that $t_{L,R}=\tan\frac12 \varphi_{L,R}$. The normalization factor $c^{j,\al}$ is easily computed in the limit $\rho\rar \infty$, where the dependences of the integrand on $t_L$ and $t_R$ factorize. We find
\bea
c^{j,\al}=\frac{4^{2j} \sin \pi 2j}{\sin\pi(j-\al)\ \sin\pi(j+\al)}\ .
\eea 

\subsection{Completeness of the bases of functions \label{secbas}}

We have been considering functions on a space $X$ with $X\in\{\Hp,\SLR,AdS_3\}$. Given $X$, let us consider the space of complex-valued square-integrable functions $L^2(X)$ with the scalar product $\langle f,g\rangle=\int_X d\mu\ \bar{f}g$, where the invariant measure can be written in all three cases as $d\mu=\sinh 2\rho\ d\rho\ d\theta\ d\tau$. Although our functions $\Phi$ do not necessarily belong to $L^2(X)$, they form orthogonal bases in the same sense as $\{e^{ipq}|p\in \R\}$ is an orthogonal basis of the space of functions on $\R$. Namely, there exist sets $B_X$ of values of the parameters 
and $\{\Phi_b,b\in B_X\}$ of the corresponding functions such that any pair $(f,g)$ of smooth, compactly supported functions on $X$ obeys $\la f,g\ra = \sum_{b\in B_X} N(b) \la f,\Phi_b\ra \la \Phi_b,g\ra$, where $N(b)$ is a normalization factor, and the sum $\sum_{b\in B_X}$ becomes an integral whenever it involves continuous parameters. 

More specifically, the $x$-bases of functions are 
\bea
\Hp :&& \{\Phi^j_x|j\in -\tfrac12+i\R_+,x\in \C\}\ ,
\\
\SLR :&& \{\Phi^{j,\eta}_{t_L,t_R}|j\in -\tfrac12+i\R_+,(t_L,t_R)\in \R^2,\eta\in \{0,\tfrac12\}\}
\nn
\\ 
   &\cup& \{\Phi^{j,\eta}_{t_L,t_R}|j\in -1-\tfrac12\N,(t_L,t_R)\in \R^2,\eta=j\mod 1\}\ ,
\\
AdS_3 :&& \{\Phi^{j,\al}_{t_L,t_R}|j\in -\tfrac12+i\R_+,(t_L,t_R)\in \R^2,\al\in [0,1)\} 
\nn
\\ &
\cup &\{\Phi^{j,\al}_{t_L,t_R}|j\in (-\tfrac12,\infty),(t_L,t_R)\in \R^2,\al=j\mod 1\}\ ,
\eea
and the corresponding $m$-bases are 
\bea
\Hp : && \{\Phi^j_{m,\bar{m}}|j\in -\tfrac12+i\R_+,m+\bar{m}\in i\R,m-\bar{m}\in\Z\}\ ,
\\
\SLR :&& \{\Phi^{j}_{m,\bar{m}}|j\in -\tfrac12+i\R_+,m+\bar{m}\in \tfrac12\Z,m-\bar{m}\in\Z\}
\nn
\\ &\cup& \{\Phi^{j}_{m,\bar{m}}|j\in -1-\tfrac12\N,m,\bar{m}\in \pm(j+1+\N)\}\ ,
\\
AdS_3 :&& \{\Phi^{j}_{m,\bar{m}}|j\in -\tfrac12+i\R_+,m+\bar{m}\in \R, m-\bar{m}\in\Z\} 
\nn
\\ &
\cup &\{\Phi^{j}_{m,\bar{m}}|j\in (-\tfrac12,\infty),m,\bar{m}\in \pm(j+1+\N)\}\ .
\eea
The completeness of both the $x$- and $m$-bases of functions on $\Hp$ was proved in \cite{tes97b}. In the case of $AdS_3$, the completeness of the $m$-basis follows from the results of \cite{bas07}, where a Plancherel formula for $AdS_3$ was proved. The completeness of the $t$-basis then follows from the integral relation (\ref{pmpt}). The case of $\SLR$ can be deduced from the case of $AdS_3$ by noting that functions on $\SLR$ correspond to $\tau$-periodic functions on $AdS_3$ with period $2\pi$. 

Moreover, in each case the basis $\{\Phi_b,b\in B_X\}$ provides a spectral decomposition of the Laplacian on $X$, which is Hermitian with respect to the scalar product $\langle f,g\rangle$. A function of spin $j$ is an eigenvector of the Laplacian for the eigenvalue $-j(j+1)$. In the cases $X\in\{\SLR,AdS_3\}$ this follows from the transformation properties of such functions under the symmetries, and the fact that the Laplacian coincides with the Casimir differential operators associated with these symmetries. In the case of $\Hp$ this can be deduced from the case of $AdS_3$ by Wick rotation.

\zeq\section{Representations and Clebsch-Gordan coefficients}

\subsection{Representations of $s\ell(2,\R)$}

The minisuperspace limit of the spectrum of the $AdS_3$ WZNW model is the space of delta-function normalizable functions on $AdS_3$. It is subject to the action of the geometrical symmetry group $\frac{AdS_3\times AdS_3}{\Z}$, and therefore of its Lie algebra $s\ell(2,\R)\times s\ell(2,\R)$. Three types of unitary representations of $s\ell(2,\R)$ appear in the minisuperspace spectrum: continuous representations, and two series of discrete representations. Continuous representations $C^{j,\al}$ are parametrized by a spin $j$ and a number $\al\in [0,1)$ such that $m\in \al +\Z$. Discrete representations $D^{j,\pm}$ are parametrized by a spin $j\in (-\frac12,\infty)$, and their states obey $m\in \pm (j+1+\N)$. All these representations of $s\ell(2,\R)$ extend to representations of the group $AdS_3$. However, only representations with $m\in \frac12\Z$, namely $C^{j,\al}$ with $\al \in \frac12\Z$ and $D^{j,\pm}$ with $j\in \frac12 \N$, extend to representations of the group $\SLR$. 

More precisely, given the $s\ell(2,\R)$ algebra with generators $J^3,J^\pm$ and relations $[J^3,J^\pm]=\pm J^\pm,\ [J^+,J^-]=-2J^3$, the spin $j$ is defined by $(J^3)^2-\frac12(J^+J^-+J^-J^+)=j(j+1)$, and the states $|m\rangle$ are such that
\bea
J^3|m\rangle = m|m\rangle \scs J^+|m\rangle = (m+j+1) |m+1\rangle \scs J^-|m\rangle =(m-j-1)|m-1\rangle\ .
\label{jjj}
\eea
These conventions are incompatible with the unit normalization of the states (which would mean $\langle m|m'\rangle = \delta_{m,m'}$), however they will turn out to agree with the behaviour of our functions $\Phi^j_{m,\bar{m}}$ eq. (\ref{pjh}). 

The tensor product laws for $s\ell(2,\R)$ representations are well-known. They are equivalent to knowing the three-point invariants, which we schematically depict here in the cases when they do not vanish: 
\bea
\psset{unit=.5cm}
\pspicture[](-2,-2)(3,2.5) 
\psline(0,0)(0,2) \rput{120}(0,0){\psline(0,0)(0,2)} \rput{-120}(0,0){\psline(0,0)(0,2)} 
\rput[t](0,-.3){$2$}
\rput[t](0,-1.5){$C\otimes C\otimes C$}
\endpspicture
,
\pspicture[](-3,-2)(3,2) 
\psline(0,0)(0,2) \rput{120}(0,0){
\psline(0,0)(0,2) \psline[arrowsize=2pt 4,arrowinset=.6]{->}(0,0)(0,1.4)} \rput{-120}(0,0){\psline(0,0)(0,2)} 
\rput[t](0,-1.5){$D^-\otimes C\otimes C$}
\endpspicture
,
\pspicture[](-3,-2)(3,2) 
\psline(0,0)(0,2) \rput{120}(0,0){\psline(0,0)(0,2) \psline[arrowsize=2pt 4,arrowinset=.6]{->}(0,2)(0,.6)} \rput{-120}(0,0){\psline(0,0)(0,2) \psline[arrowsize=2pt 4,arrowinset=.6]{->}(0,0)(0,1.4)} 
\rput[t](0,-1.5){$D^-\otimes D^+\otimes C$}
\endpspicture
,
\pspicture[](-3,-2)(2,2) 
\psline(0,0)(0,2)\psline[arrowsize=2pt 4,arrowinset=.6]{->}(0,2)(0,.6) 
 \rput{120}(0,0){\psline(0,0)(0,2) \psline[arrowsize=2pt 4,arrowinset=.6]{->}(0,2)(0,.6)} \rput{-120}(0,0){\psline(0,0)(0,2) \psline[arrowsize=2pt 4,arrowinset=.6]{->}(0,0)(0,1.4)} 
\rput[t](0,-1.5){$D^+\otimes D^+\otimes D^-$}
\endpspicture
.
\label{tpl}
\eea
For instance, the first diagram means that any continuous representation $C^{j,\al}$ appears twice in the tensor product $C^{j_1,\al_1}\otimes C^{j_2,\al_2}$ of two continuous representations. (The $m$-conservation rule $\al=\al_1+\al_2 \mod 1$ is implicitly assumed.) The fourth diagram means that $D^{j,+} \subset D^{j_1,+}\otimes D^{j_2,+}$. (The rule $j\in j_1+j_2+1+\N$ is implicitly assumed.) The fourth diagram also means that $D^{j,-}$ may appear once in $D^{j_1,-}\otimes D^{j_2,+}$. (This happens if $j\in j_2-j_1-1-\N$.) We omit the diagrams obtained by reverting the arrows in the second and fourth diagrams, namely $D^+\otimes C\otimes C$ and $D^-\otimes D^-\otimes D^+$.

\subsection{Clebsch-Gordan coefficients: $m$ basis}

We will rederive the tensor product rules by studying the Clebsch-Gordan coefficients. These coefficients are the three-point invariants, viewed as functions $C(j_1,j_2,j_3|m_1,m_2,m_3)$ subject to the equations 
\bea
\sum_{i=1}^3 (m_i+j_i+1) C(m_i+1) =
\sum_{i=1}^3 (m_i-j_i-1) C(m_i-1) =
\sum_{i=1}^3 m_i C = 0 \ .
\label{cdef}
\eea
It is of course possible to prove a priori that these equations are obeyed by the three-point function $\la \prod_{i=1}^3 \Phi^{j_i}_{m_i,\bar{m}_i}\ra \equiv \int_{AdS_3} dG \prod_{i=1}^3 \Phi^{j_i}_{m_i,\bar{m}_i}(G)$. To do this, we would introduce a realization of the Lie algebra $s\ell(2,\R)$ as first-order differential operators $D^a$ wrt $\rho,\theta,\tau$, such that $D^+ \Phi^j_{m,\bar{m}}(G) = (m+j+1) \Phi^j_{m,\bar{m}}(G),\ D^-\Phi^j_{m,\bar{m}}(G)=(m-j-1) \Phi^j_{m,\bar{m}}(G)$ and $D^3 \Phi^j_{m,\bar{m}}(G) = m \Phi^j_{m,\bar{m}}(G)$. Then eq. (\ref{cdef}) would follow from the identity $\la \prod_{i=1}^3 \Phi^{j_i}_{m_i,\bar{m}_i}(G)\ra = \la \prod_{i=1}^3 \Phi^{j_i}_{m_i,\bar{m}_i}(G_LG)\ra$. We however abstain from doing this, as we will later explicitly compute the three-point function $\la \prod_{i=1}^3 \Phi^{j_i}_{m_i,\bar{m}_i}\ra$ and write it in terms of solutions of the equation (\ref{cdef}). 

Given three irreducible representations of $s\ell(2,\R)$, there exist zero, one or two linearly independent solutions of the equation (\ref{cdef}). In the case of three continuous representations, the momenta $(m_1,m_2)$ belong to a two-dimensional lattice of the type $\prod_{i=1}^2(\al_i + \Z)$ (with of course $m_3=-m_1-m_2$), and the coefficients $m_i\pm (j_i+1)$ never vanish. In this situation, a solution of eq. (\ref{cdef}) is determined once the values of $C$ at two neighbouring points of the lattice are given. In the case when at least one representation is discrete, say $m_1\in j_1+1+\N$, the lattice becomes semi-infinite in one direction, and a solution is determined once the value of $C$ at one point is given. 

Let us introduce the function
\bea
\tgt{a}{b}{c}{e}{f}\equiv\frac{\G(a)\G(b)\G(c)}{\G(e)\G(f)}\tft{a}{b}{c}{e}{f}{1} = \sum_{n=0}^\infty \frac{1}{n!} \frac{\G(a+n)\G(b+n)\G(c+n)}{\G(e+n)\G(f+n)}\ ,
\label{gf}
\eea
where the sum converges provided $a+b+c-e-f<0$, and the poles of $G$ are the same as those of $\G(a)\G(b)\G(c)\G(e+f-a-b-c)$. Knowing the identity
\bea
(a-e+1) \tgt{a}{b}{c}{e}{f} + (b-f)\tgt{a+1}{b}{c}{e}{f+1} + (c-1)\tgt{a}{b}{c-1}{e-1}{f}=0\ ,
\label{tgtp}
\eea
we can use this function for writing a solutions of eq. (\ref{cdef}):
\bea
C=\delta(m_1+m_2+m_3)\ \tgt{-j_2+m_2}{-j_3-m_3}{-j_{23}^1}{1+j_1-j_3+m_2}{1+j_1-j_2-m_3}  \equiv \delta(m_1+m_2+m_3) g^{23}\ ,
\label{cgg}
\eea
which is well-defined provided $2+j_{123}> 0$, where we use the notations $j_{123}=j_1+j_2+j_3$ and
$j_{23}^1\equiv j_2+j_3-j_1$. 
Of course five other solutions of the type $g^{ab}$ with $a\neq b\in \{1,2,3\}$ can be obtained by permutations of indices. These solutions are not linearly independent, as can be shown with the help of the identity
 \begin{multline}
s(b)s(c-a) \tgt{a}{b}{c}{e}{f}=
s(e-a)s(f-a)\tgt{a}{a-e+1}{a-f+1}{a-b+1}{a-c+1}
\\
-s(c-e)s(c-f)\tgt{c}{c-e+1}{c-f+1}{c-a+1}{c-b+1}\ ,
\end{multline}
where $s(x)\equiv \sin \pi x$, and we will also use $c(x)\equiv \cos \pi x$. Thus we obtain
\bea
\left(\begin{array}{c} g^{21} \\ g^{12} \end{array}\right) = M_{13}\left(\begin{array}{c} g^{32} \\ g^{23} \end{array}\right) \scs M_{13}=\frac{1}{s(j_{12}^3)}\bsm \frac{s(j_2+m_2) s(j_3-m_3)}{s(j_1+m_1)} & \frac{s(j_3+m_3)s(j_3-j_1+m_2)}{s(j_1+m_1)} \\ \frac{s(j_3-m_3)s(j_3-j_1-m_2)}{s(j_1-m_1)} & \frac{s(j_2-m_2)s(j_3+m_3)}{s(j_1-m_1)} \esm \ .
\label{ggmgg}
\eea
Together with the other identities obtained by permuting the indices, this shows 
that at most two of the solutions $g^{ab}$ are linearly independent.

Due to our convention $j\in (-\frac12,\infty)$ for discrete representations, we have $\Re j\geq -\frac12$ for all representations of interest. This ensures that the sum in eq. (\ref{gf}) converges, so that $g^{ab}$ is well-defined provided the summand is finite, which occurs unless $\G(-j_a+m_a)\G(-j_b-m_b)\G(-j_{ab}^c)$ has a pole.

\paragraph{Case $C\otimes C\otimes C$.} 

In this case, two given solutions say $g^{23},g^{32}$ are linearly independent, and they provide a basis of the two-dimensional space of invariants.

\paragraph{Case $D^-\otimes C\otimes C$.} 

We assume for example $m_2= -j_2-1-\ell$ with $\ell\in \N$. Some relations of the type of eq. (\ref{ggmgg}) simplify, and we find
\bea
g^{21}=(-1)^\ell \frac{s(j_3+m_3)}{s(j_1+m_1)} g^{23} = -\frac{s(j_{13}^2)}{s(2j_2)}g^{31} = -(-1)^\ell \frac{s(j_3+m_3)}{s(j_1+m_1)} \frac{s(j_{13}^2)}{s(2j_2)} g^{13} \ .
\label{ggggm}
\eea
Since the space of invariants is one-dimensional, the two remaining functions $g^{12}$ and $g^{32}$ must also be proportional to the other four. However, this proportionality relation is not very simple, as can be seen in the case of the highest-weight state $\ell=0$ when $g^{12}$ and $g^{32}$ fail to become expressible as products of $\G$-functions, in contrast to the other four solutions.

\paragraph{Case $D^+\otimes C\otimes C$.}

The situation is completely analogous to the previous case. We assume for example $m_2=j_2+1+\ell$ with $\ell\in\N$ and find
\bea
g^{12}=(-1)^\ell \frac{s(j_3-m_3)}{s(j_1-m_1)} g^{32} = -\frac{s(j_{13}^2)}{s(2j_2)} g^{13} = -(-1)^\ell \frac{s(j_3-m_3)}{s(j_1-m_1)}\frac{s(j_{13}^2)}{s(2j_2)} g^{31}\ .
\label{ggggp}
\eea

\paragraph{Case $D^-\otimes D^-\otimes C$.}

We expect no invariants to exist in this case. Let us check this, assuming for example $m_2\in -j_2-1-\ell_2$ and $m_3\in -j_3-1-\ell_3$ with $\ell_2,\ell_3\in \N$.
Equation (\ref{ggggm}) implies two incompatible relations between $g^{21}$ and $g^{31}$, which must therefore both vanish. An apparent paradox comes from the non-vanishing of $g^{12},g^{32},g^{23},g^{13}$. However, these functions do not provide solutions to eq. (\ref{cdef}), because they become infinite at $\ell_2=-1$ or $\ell_3=-1$. For instance, if $C(m_1,-j_2,m_3)=\infty$, then $(m_1+j_1+1)C(m_1+1,-j_2-1,m_3)+ (-j_2+j_2)C(m_1,-j_2,m_3) + (m_3+j_3+1)C(m_1,-j_2-1,m_3+1) =0$ may have an unwanted nonvanishing second term. 
Therefore, the analysis of $g^{ab}$ agrees with the representation-theoretic expectations that no invariant exists.

\paragraph{Case $D^+\otimes D^-\otimes C$.} 

We assume for example $m_2=-j_2-1-\ell_2$ and $m_3=j_3+1+\ell_3$ with $\ell_2,\ell_3\in\N$. We find the relations
\bea
g^{31}=g^{12}=-\frac{s(2j_3)}{s(j_{12}^3)} g^{13}=-\frac{s(2j_2)}{s(j_{13}^2)} g^{21} = \frac{s(2j_2)s(2j_3)}{s(j_{13}^2) s(j_{12}^3)} g^{23}\ .
\label{mpg}
\eea
The functions $g^{13},g^{21},g^{23}$ stay finite for any values $\ell_2,\ell_3\in\Z$, and therefore provide three proportional invariants. The functions $g^{31}$ and $g^{12}$ become infinite if $\ell_3<0$ and $\ell_2<0$ respectively, so that it is not a priori clear that they provide invariants. That they actually do is guaranteed by the above relations. 

\paragraph{Case $D^-\otimes D^-\otimes D^+$.} 

We assume for example $m_1= j_1+1+\ell_1$, $m_2\in -j_2-1-\ell_2$ and $m_3\in -j_3-1-\ell_3$, with $\ell_1,\ell_2,\ell_3\in \N$. Noticing $s(2j_2)s(2j_3)=s(j_{13}^2)s(j_{12}^3)$, we find the relations
\bea
g^{12}=g^{23}=-\frac{s(2j_3)}{s(j_{12}^3)} g^{13}=-\frac{s(2j_3)}{s(j_{12}^3)} g^{32}\ .
\eea
These four proportional functions provide the invariant in this case. In particular, $g^{12}$ and $g^{13}$ no longer become infinite at $\ell_2=-1$ or $\ell_3=-1$ respectively, as happened in the case $D^-\otimes D^-\otimes C$. The remaining two functions $g^{21},g^{31}$ vanish, as they already did in the case $D^-\otimes D^-\otimes C$. Notice that the selection rule $j_1\in j_2+j_3+1+\N$ manifests itself as $g^{23}$ becoming infinite if $j_1\in j_2+j_3-\N$, due to a series of poles which correspond to those of $\G(-j_{23}^1)$.

\subsection{Clebsch-Gordan coefficients: $t$-basis}

Our $m$-basis invariants $g^{ab}$ are not symmetric under permutations of the indices, but the equation (\ref{cdef}) which they solve is. In the $t$-basis, we will now show that there exist natural permutation-symmetric invariants, and we will relate them to combinations of the $g^{ab}$ invariants. A similar analysis was already performed for the Clebsch-Gordan coefficients of $SO(2,1)=\frac{\SLR}{\Z_2}$, in the articles \cite{kv97,kv98}. The representations of $SO(2,1)$ correspond to representations of $s\ell(2,\R)$ such that $m\in \Z$, or in other words $\al=0$. We will perform the generalization to arbitrary values of $\al$.   

In the $t$-basis, $AdS_3$ invariants should be solutions of 
\bea
C(\{t_i\})=C(\{\tfrac{at_i+b}{ct_i+d}\})\prod_{i=1}^3 |ct_i+d|^{2j_i} e^{-2i\pi\al_iN(G|t_i)}\ ,
\eea 
for any $AdS_3$ element $G$ whose projection onto $\SLR$ is $g=\bsm a & b \\ c & d\esm$. 
Solutions exist provided $\al_1+\al_2+\al_3\in \Z$, and we will assume $\al_1+\al_2+\al_3=0$.
The solutions are \cite{rib07}
\bea
C(\{t_i\})=|t_{12}|^{j_{12}^3} |t_{23}|^{j_{23}^1} |t_{31}|^{j_{31}^2}\  e^{i\pi\left(\al_{12}\sign t_{12}+\al_{23}\sign t_{23} +\al_{31}\sign t_{31}\right)}\ ,
\label{ct}
\eea
where $\al_{12},\al_{23},\al_{31}$ should obey the equation $\al_{12}-\al_{23}=\al_2 \mod \Z$ and the two other equations obtained by even permutations thereof. The solutions to such equations are 
\bea
\al_{ab}=\tfrac13(\al_b-\al_a) +\al_0\ ,
\label{alz}
\eea
where $\al_0$ is an arbitrary constant. (In $\SLR$ we have $\al_a=\eta_a\in \frac12\Z$ and it is more convenient to adopt the convention $\al_{ab}=\eta_a+\eta_b+\al_0$. Equivalently, we can use the above formula provided we assume $\al_a\in \frac32\Z$.) 
Let us now apply the change of basis (\ref{pmpt}) to $C(\{t_i\})$. After the change of integration variables $t_i=\tan\frac{\varphi_i}{2}$ the $m$-basis version of $C(\{t_i\})$ is 
\begin{multline}
C(\{m_i\}) = \prod_{i=1}^3\left[\int_{-\pi}^\pi d\varphi_i\ e^{im_i\varphi_i}\right] |\sin\tfrac12 \varphi_{12}|^{j_{12}^3} |\sin\tfrac12\varphi_{23}|^{j_{23}^1} |\sin\tfrac12 \varphi_{31}|^{j_{13}^2} 
\\ \times e^{i\pi \left(\al_{12}\sign \sin \frac12\varphi_{12} +\al_{23}\sign\sin\frac12\varphi_{23}+\al_{31}\sign\sin\frac12\varphi_{31}\right)}\ ,
\end{multline}
where the integrand has the same values at $\varphi_i=\pi$ and $\varphi_i=-\pi$. 
This integral can be explicitly evaluated by generalizing the computations of \cite{kv97}, and in particular using the formula, valid for $\varphi\in (-\pi,\pi)$:
\bea
\left|\sin \tfrac{\varphi}{2}\right|^{-2a} e^{i\pi \al \sign \sin\frac12\varphi} =s(a+\al)e^{-i\pi \al} 2^{2a}\G(1-2a)\sum_{n=-\infty}^\infty e^{i(n+\al)\varphi} \frac{\G(a+n+\al)}{\G(1-a+n+\al)} \ .
\eea
Thus, we find
\bea
C(\{m_i\})=-\delta(\tsum m_i)   2^{-j_{123}}\tfrac{1}{\pi^2} \G(1+j_{12}^3)\G(1+j_{23}^1)\G(1+j_{31}^2)
\ e^{-i\pi(\al_{12}+\al_{23}+\al_{31})}\
 g^{\al_0} \ ,
\eea
where we introduce the invariants
\begin{multline}
g^{\al_0} = \pi^2 s(\tfrac12 j_{12}^3 - \al_{12}) s(\tfrac12 j_{23}^1 -\al_{23}) s(\tfrac12 j_{31}^2 -\al_{31})
 \\ \times
 \sum_{n\in \Z} 
\frac{\G(-\frac12 j_{12}^3     +\al_{12}+n)}{\G(1+\frac12 j_{12}^3+    \al_{12}+n)} 
\frac{\G(-\frac12 j_{13}^2 +m_1+\al_{12}+n)}{\G(1+\frac12 j_{13}^2+m_1+\al_{12}+n)} 
\frac{\G(-\frac12 j_{23}^1 -m_2+\al_{12}+n)}{\G(1+\frac12 j_{23}^1-m_2+\al_{12}+n)}\ ,
\end{multline}
which can be expressed in terms of the invariants $g^{ab}$ (\ref{cgg}) as
\bea
g^{\al_0}=
 s(\tfrac12 j_{13}^2-\al_{31}) s(j_3-\al_3)s(j_1+\al_1) g^{31}+s(\tfrac12 j_{13}^2+\al_{31}) s(j_1-\al_1)s(j_3+\al_3) g^{13}\ .
\label{ge}
\eea
In this formula, $g^{\al_0}$ depends on the paramter $\al_0$ only through $\al_{31}$ eq. (\ref{alz}). There are two particularly interesting special values of $\al_0$, namely $0$ and $\frac12$. In the case $\al_0=0$ then $C(\{t_i\})$ and $g^0$ are invariant under permutations. In the case $\al_0=\frac12$ then $C(\{t_i\})$ and $g^\frac12$ are odd under permutations i.e. invariant up to a sign. When the three involved representations are continuous, $g^0$ and $g^\frac12$ can serve as a basis of the two-dimensional space of invariants. Notice that in the case of $\SLR$ only these two values of $\al_0$ are possible.

\zeq\section{Products of functions}

In the conformal bootstrap approach to the $AdS_3$ WZNW model, all correlation functions can in principle be constructed from the knowledge of three objects: the spectrum, the two-point correlation functions on a sphere, and the operator product expansions -- or equivalently the three-point correlation functions on a sphere. We will now determine these objects in the minisuperspace limit. We first recall their definitions. Given $n$ functions $\Phi^i(G)$ on $AdS_3$, the corresponding correlation function is $\la \prod_{i=1}^n \Phi^i \ra \equiv \int dG\ \prod_{i=1}^n \Phi^i(G)$ where $dG$ is the invariant measure. If $\{\Phi^i\}_{i\in S}$ form an orthogonal basis of the spectrum (that is $\la \Phi^i\Phi^j\ra =0$ if $i\neq j$), the product of functions is schematically  $\Phi^1\Phi^2=\sum_{i\in S} 
\frac{\la \Phi^1\Phi^2\Phi^i \ra}{\la\Phi^i\Phi^i\ra} \Phi^i $. This product is obviously associative and commutative.

The functions $\Phi^j_{m,\bar{m}}(G)$ on $AdS_3$ are related to corresponding functions $\Phi^j_{m,\bar{m}}(h)$ on $\Hp$ by a Wick rotation, and therefore their correlation functions can be deduced from $\Hp$ correlation functions by that Wick rotation. This will involve some subtleties, because the discrete representations which appear in the minisuperspace spectrum on $AdS_3$ are absent in $\Hp$. But let us first review the products of functions on $\Hp$.

\subsection{Products of functions on $\Hp$}

The minisuperspace spectrum of $\Hp$ is generated by the functions
\bea
\{\Phi^j_{n,p}(h)|j\in -\tfrac12+i\R,\ n\in\Z,\ p\in i\R\}\ .
\eea
The correlation functions $\la \prod_{i=1}^n \Phi^{j_i}_{n_i,p_i} \ra \equiv \int dh\ \prod_{i=1}^n \Phi^{j_i}_{n_i,p_i}(h)$ are obtained by integrating products of such functions with respect to the invariant measure
$dh=\sinh 2\rho\ d\rho\ d\theta\ d\tau$. The two-point functions can be computed from the expression (\ref{pjh}) of $\Phi^j_{n,p}(h)$:
\bea
\la \Phi^j_{n,p} \Phi^{j'}_{n',p'}\ra &= & 128 \pi^2 \delta_{n+n',0}\ \delta(p+p') \left[ \delta(j+j'+1) + R^j_{n,p}\delta(j-j')\right]\ ,
\label{pjpj}
\eea
where the reflection coefficient $R^j_{n,p}$ was defined in eq. (\ref{rj}).
A similar direct computation of the three-point function seems complicated. Instead, we will make use of the known $x$-basis three-point function \cite{tes97b}
\bea
\la \prod_{i=1}^3 \Phi^{j_i}_{x_i} \ra &=& C(j_1,j_2,j_3) |x_{12}|^{2j_{12}^3} |x_{23}|^{2j_{23}^1} |x_{31}|^{2j_{31}^2}\ ,
\label{xxx}
\\
C(j_1,j_2,j_3) &\equiv & \pi^{-3} \G(-j_{123}-1)\frac{\G(-j_{12}^3)\G(-j_{23}^1)\G(-j_{31}^2)}{\G(-2j_1-1)\G(-2j_2-1)\G(-2j_3-1)}\ .
\label{cfor}
\eea
The transformation to the $m$-basis (\ref{phph}) can be performed thanks to an integral formula of Fukuda and Hosomichi \cite{fh01}. The result can be written in terms of the Clebsch-Gordan coefficients $g^{ab}$ (\ref{cgg}):
\begin{multline}
\la \prod_{i=1}^3 \Phi^{j_i}_{m_i,\bar{m}_i} \ra = C(j_1,j_2,j_3)\ \delta^{(2)}(\tsum m_i)\ K^2s(j_{13}^2)^2 s(j_{23}^1)^2
\\ \times
\left[g^{23}\bar{g}^{31} +\frac{s(2j_2)}{s(j_{13}^2)} g^{23}\bar{g}^{32} +\frac{s(2j_1)}{s(j_{23}^1)}g^{13}\bar{g}^{31} +g^{13}\bar{g}^{32}\right]\ ,
\label{ppp}
\end{multline}
where $\bar{g}^{ab}$ denotes $g^{ab}$ with $m_i$ replaced by $\bar{m}_i$, and we introduced the factor
\bea
K\equiv \frac{1}{\pi ^2}\G(1+j_{12}^3)\G(1+j_{13}^2)\G(1+j_{23}^1)\ .
\label{kfor}
\eea
It can be checked that the two- and three-point functions have the behaviour under reflection which is expected from the behaviour of $\Phi^j_{m,\bar{m}}$ (\ref{rj}). 
Now, using the three-point function, products of functions on $\Hp$ can be written as 
\bea
\Phi^{j_1}_{m_1,\bar{m}_1}\Phi^{j_2}_{m_2,\bar{m}_2} = \frac{1}{256\pi^2}\int_{-\frac12+i\R} dj_3 \ \la \prod_{i=1}^3 \Phi^{j_i}_{m_i,\bar{m}_i} \ra' \frac{1}{R^{j_3}_{m_3,\bar{m}_3}} \Phi^{j_3}_{-m_3,-\bar{m}_3}\ ,
\label{php}
\eea
where we use the notation $\la \prod_{i=1}^3 \Phi^{j_i}_{m_i,\bar{m}_i} \ra = \delta^{(2)}(m_1+m_2+m_3)\la \prod_{i=1}^3 \Phi^{j_i}_{m_i,\bar{m}_i} \ra' $.

The invariance of the three-point function (\ref{ppp}) under permutations of the indices is not manifest, but can be checked using linear relations between the $g^{ab}$ such as eq. (\ref{ggmgg}). It seems that a reasonably simple, manifestly permutation-symmetric expression exists only in the case $m_i,\bar{m}_i\in \eta_i+\Z$ with $\eta_i\in \{0,\frac12\}$, which corresponds to functions on $\SLR$. In this case, we can use the invariants $g^0,g^\frac12$ (\ref{ge}), and we find 
\begin{multline}
\la \prod_{i=1}^3 \Phi^{j_i}_{m_i,\bar{m}_i} \ra = -\tfrac{1}{2\pi^5} C(j_1,j_2,j_3)\ \delta_{\sum m_i,0} \delta_{\sum\bar{m}_i,0}\ \frac{\g(1+j_{12}^3)\g(1+j_{23}^1)\g(1+j_{31}^2)}{\prod_{i=1}^3 s(j_i+\eta_i)}
\\
\times \sum_{\e\in \{0,\frac12\}} \frac{s(\tfrac12 j_{123}+\e)}{ s(\tfrac12 j_{12}^3+\eta_3+\e) s(\tfrac12 j_{23}^1 +\eta_1+\e) s(\tfrac12 j_{31}^2 +\eta_2 +\e)} \ g^\e \bar{g}^\e\ ,
\label{gbg}
\end{multline}
where we use $\g(x)\equiv \frac{\G(x)}{\G(1-x)}$. From this, we can reconstruct the $t$-basis three-point function
\begin{multline}
\la \prod_{i=1}^3 \Phi^{j_i,\eta_i}_{t^L_i,t^R_i}\ra = \frac{2^{2j_{123}}}{2\pi^4} C(j_1,j_2,j_3) \prod_{i=1}^3 c(j_i-\eta_i)\ \times 
\\
\left|t_{12}^L t_{12}^R\right|^{j_{12}^3} \left|t_{23}^L t_{23}^R\right|^{j_{23}^1} \left|t_{31}^Lt_{31}^R\right|^{j_{31}^2}
e^{i\pi\left[\eta_3 (\sign t_{12}^L +\sign t_{12}^R) +\eta_1 (\sign t_{23}^L +\sign t_{23}^R) +\eta_2( \sign t_{31}^L+\sign t_{31}^R)\right]} \times
 \\ 
\sum_{\e\in \{0,\frac12\}} s(\tfrac12 j_{123}-\e) c(\tfrac12 j_{12}^3+\eta_3+\e) c(\tfrac12 j_{23}^1 +\eta_1+\e) c(\tfrac12 j_{31}^2 +\eta_2 +\e)\ \left(\sign t_{12}^Lt_{12}^Rt_{23}^Lt_{23}^R t_{31}^Lt_{31}^R\right)^{2\e}\ .
\end{multline}
Comparing this formula to the $\Hp$ three-point function in the $x$-basis eq. (\ref{xxx}), we obtain a confirmation of the lack of a simple relation between the $x$-basis in $\Hp$ and the $t$-basis in $\SLR$ or $AdS_3$. 

In the more general case of functions on $AdS_3$, the three-point function can still be expressed in terms of the invariants $g^0$ and $g^{\frac12}$, but the formula is more complicated than eq. (\ref{gbg}) and in particular the ``mixed'' terms $g^0\bar{g}^\frac12$ and $g^\frac12 \bar{g}^0$ are present. Their absence in the case of $\SLR$ can be attributed to the exterior automorphism $\omega$ of $\SLR$, namely $\omega(g)=\bsm 1 & 0 \\ 0 & -1 \esm g \bsm 1 & 0 \\ 0 & -1 \esm$, which is such that $\Phi^{j,\eta}_{t_L,t_R}(\omega(g)) = (-1)^{2\eta} \Phi^{j,\eta}_{-t_L,-t_R}(g)$. In the case of $AdS_3$ this action still exists and can be expressed as $\omega(\rho,\theta,\tau)=(\rho,-\theta,-\tau)$. But it does not act simply on the function $\Phi^{j,\al}_{t_L,t_R}(G)$.\footnote{Let us give the behaviour of certain objects of section \ref{ssect}: $N(\omega(G)|t)=-N(G|-t)$, $[\omega(G)]=-[G]-1$ and $n(\omega(G)|t_L,t_R)=-n(G|-t_L,-t_R)$.}

\subsection{Products of functions on $AdS_3$}

Functions $\Phi^j_{m,\bar{m}}$ on $AdS_3$ are obtained from the corresponding functions on $\Hp$ by performing the Wick rotation $\tau\rar i\tau$ and continuing $p=m+\bar{m}$ from $i\R$ to $\R$. If we do not modify the value of the spin $j\in -\frac12+i\R$, this yields functions transforming in the continuous representations of $AdS_3\times AdS_3$, namely $\Phi^j_{m,\bar{m}} \in C^{j,\al}\otimes C^{j,\al}$ where $m,\bar{m}\in \al+\Z$. We may in addition obtain functions transforming in the discrete representation by continuing $j$ to real values such that $m,\bar{m}\in \pm j \pm \Z$. More precisely, functions $\Phi^j_{m,\bar{m}}\in D^{j,\pm}\otimes D^{j,\pm}$ correspond to 
\bea
 j\in (-\tfrac12,\infty) \qquad &{\rm and}& \qquad  m,\bar{m}\in \pm(j+1+\N) \Leftrightarrow  -j-1-\tfrac12|n|\pm \tfrac12 p \in \N\ ,
 \label{jpos}
 \\
 {\rm or} \qquad j\in (-\infty,-\tfrac12) \qquad &{\rm and}& \qquad m,\bar{m}\in \pm(-j+\N) \Leftrightarrow j-\tfrac12|n|\pm \tfrac12 p \in \N\ .
 \label{jneg}
\eea
These two possibilities are related by the reflection $j\rar -j-1$ and they are equivalent. We will only consider the first possibility, because our invariants $g^{ab}$ (\ref{cgg}) are well-defined for $\Re j\geq -\frac12$. 
We will see that 
the set of these discrete and continuous functions is closed under products, consistently with the fact that they
generate the space of functions on $AdS_3$ as we saw in section \ref{secbas}. 

We now derive the products of functions on $AdS_3$ by continuing the products of functions on $\Hp$ (\ref{php}) to the relevant values of spins $j$ and momenta $p$. 
We will examine various cases, according to the nature -- discrete or continuous -- of the fields $\Phi^{j_1}_{m_1,\bar{m}_1}$ and $\Phi^{j_2}_{m_2,\bar{m}_2}$. For example, the case when $j_1\in -\frac12+i\R$ and $m_2,\bar{m}_2\in j_2+1+\Z$ will be denoted $C\times D^+$. We will check that the terms which appear in a given product are those which are allowed by the well-known tensor product laws for $s\ell(2,\R)$ representations (\ref{tpl}). 

\paragraph{Case $C\times C$.} 

We should continue $p_1,p_2,p_3$ from imaginary to real values in eq. (\ref{php}). This is problematic only when the integrand, viewed as a function of $j_3$, has poles which cross the integration line. Such $p_i$-dependent poles of the integrand may come from either of its three factors.
The poles coming from the second factor  $\frac{1}{R^{j_3}_{m_3,\bar{m}_3}}$ are easily seen from eq. (\ref{rj}), and the poles from the other factors are obtained from these by the reflection\footnote{It is also possible to study the $m$-dependent poles of $\la\prod_{i=1}^3 \Phi^{j_i}_{m_i,\bar{m}_i} \ra$ directly from the formula (\ref{ppp}). For example, $\bar{g}^{32}$ has poles at $j_2+\bar{m}_2\in\N$. But the coefficient of $\bar{g}^{32}$ is a combination of $g^{13}$ and $g^{23}$ of the type $s(2j_2)g^{23}+s(j_{13}^2)g^{13}=\frac{s(j_3-m_3)}{s(j_1-m_1)} s(j_2+m_2) g^{32}$, where we used $j_2+m_2\in\Z$ (which follows from $j_2+\bar{m}_2\in\N$) and eq. (\ref{ggmgg}). This vanishes, unless $g^{32}$ itself has a pole. This shows that $\la\prod_{i=1}^3 \Phi^{j_i}_{m_i,\bar{m}_i} \ra$ has simple poles when both $m_2,\bar{m}_2$ belong to $-j_2+\N$, but not when only $\bar{m}_2$ does.}
$j_3\rar -j_3-1$. All these poles fall on the four dashed half-lines in the following diagram, which depicts the $j_3$ complex plane. The two half-lines on the left correspond to the poles of 
$\frac{1}{R^{j_3}_{m_3,\bar{m}_3}}$.
\bea
\psset{unit=1.1cm}
\pspicture[](-5,-2.2)(5,2.2)  
\psline{->}(0,-2)(0,2)
\psline(-5,0)(5,0)
\rput[r]{0}(-.2,2){$j_3\in -\frac12+i\R$}
\rput[l]{0}(2,1){\mypoles{3}}
\rput[t]{0}(2,.8){\footnotesize $\frac12(|n_3|+p_3)$}
\rput[r]{180}(-2,-1){\mypoles{3}}
\rput[b]{0}(-2,-.8){\footnotesize $-1-\frac12(|n_3|+p_3)$}
\rput[l]{0}(2,-1){\mypoles{3}}
\rput[b]{0}(2,-.8){\footnotesize $\frac12(|n_3|-p_3)$}
\rput[r]{180}(-2,1){\mypoles{3}}
\rput[t]{0}(-2,.8){\footnotesize $-1-\frac12(|n_3|-p_3)$}
\endpspicture 
\eea
This diagram assumes $p_3\in i\R$. When $p_3$ moves to real values, let us consider the poles from the left which may cross the integration line and end up on the right with $j_3\in (-\frac12,\infty)$. Such poles belong to the (possibly empty) sets
$j_3\in -1-\tfrac12|n_3|\pm \tfrac12 p_3 -\N \cap (-\tfrac12,\infty)$.
Therefore, according to eq. (\ref{jpos}), they correspond to functions $\Phi^{j_3}_{m_3,\bar{m}_3}$ in the $D^{j_3,\pm}$ representations. We deduce the formula for the products of two ``continuous'' functions on $AdS_3$:
\begin{multline}
\Phi^{j_1}_{m_1,\bar{m}_1}\Phi^{j_2}_{m_2,\bar{m}_2} = \frac{1}{256\pi^2}\int_{-\frac12+i\R} dj_3 \ \la \prod_{i=1}^3 \Phi^{j_i}_{m_i,\bar{m}_i} \ra' \frac{1}{R^{j_3}_{m_3,\bar{m}_3}} \Phi^{j_3}_{-m_3,-\bar{m}_3}
\\
+\frac{2}{256\pi^2} \sum_{j_3\in -1-\tfrac12|n_3|\pm \tfrac12 p_3 -\N \cap (-\tfrac12,\infty)} \la \prod_{i=1}^3 \Phi^{j_i}_{m_i,\bar{m}_i} \ra' 2\pi i \Res \frac{1}{R^{j_3}_{m_3,\bar{m}_3}} \Phi^{j_3}_{-m_3,-\bar{m}_3}\ ,
\label{adsp}
\end{multline}
where the factor $2$ in the discrete term is due to the contribution of the poles with $j_3\in (-\infty,-\frac12)$. Notice that the general expression (\ref{ppp}) for the three-point function $\la \prod_{i=1}^3 \Phi^{j_i}_{m_i,\bar{m}_i} \ra'$ simplifies in the case $j_3\in -1-\frac12|n_3|\pm \frac12 p_3 -\N \cap (-\frac12,\infty)$ due to formulas of the type of eqs. (\ref{ggggm}) and (\ref{ggggp}). Examples of simplified expressions are:
\bea
\la \prod_{i=1}^3 \Phi^{j_i}_{m_i,\bar{m}_i} \ra' \underset{D^{j_3,+}}{=} -C(j_1,j_2,j_3) K^2 s(j_{12}^3) s(j_{23}^1) s(j_{31}^2) s(j_{123})\ g^{21}\bar{g}^{13} && ,
 \label{sdp}
 \\ 
 g^{21}\bar{g}^{13}= \bar{g}^{21}g^{13}=g^{12}\bar{g}^{23}=\bar{g}^{12}g^{23} && ,
 \nn
\\
\la \prod_{i=1}^3 \Phi^{j_i}_{m_i,\bar{m}_i} \ra' \underset{D^{j_3,-}}{=} -C(j_1,j_2,j_3) K^2 s(j_{12}^3) s(j_{23}^1) s(j_{31}^2) s(j_{123})\ g^{12}\bar{g}^{31} && ,
\label{sdm}
\\ 
 g^{12}\bar{g}^{31}= \bar{g}^{12}g^{31}=g^{21}\bar{g}^{32}=\bar{g}^{21}g^{32} && .
\nn
\eea

\paragraph{Case $D^+\times C$.} 

After moving $p_i$ to real values as in the previous case $C\times C$, we should move $j_1$ to $-1-\frac12 |n_1|+\frac12 p_1 -\N \cap (-\frac12,\infty)$. Let us show that no further poles cross the integration line in eq. (\ref{adsp}) during this operation. We are looking for possible $j_1$-dependent poles in $\la \prod_{i=1}^3 \Phi^{j_i}_{m_i,\bar{m}_i} \ra'$, viewed as a function of $j_3$. 
We use formulas of the type of eq. (\ref{ggggp}) to obtain
\bea
\la \prod_{i=1}^3 \Phi^{j_i}_{m_i,\bar{m}_i} \ra' \underset{D^{j_1,+}}{=} C(j_1,j_2,j_3) K^2 s(j_{12}^3) s(j_{23}^1) s(j_{31}^2) s(j_{123}) \frac{s(j_{23}^1)}{s(2j_1)} g^{23}\bar{g}^{32}\ .
\eea
Potential poles come from factors $\G(1+j_{23}^1)$ in $K$ (\ref{kfor}) and $\G(-1-j_{123})\G(-j_{12}^3)\G(-j_{31}^2)$ in $C(j_1,j_2,j_3)$ (\ref{cfor}), but they are all cancelled by appropriate $\sin$ factors. On the other hand, the poles from the factors $\G(1+j_{12}^3)\G(1+j_{31}^2)$ in $K$, from the factor $\G(-j_{23}^1)$ in $C(j_1,j_2,j_3)$, and the poles of $\G(-j_{23}^1)$ which come from $g^{23}$ and $\bar{g}^{32}$, cannot be reached because $\Re j_1\geq -\frac12$. 

This shows that the formula (\ref{adsp}) still holds for products of functions in $D^+\times C$. Of course, simplified expressions for $\la \prod_{i=1}^3 \Phi^{j_i}_{m_i,\bar{m}_i} \ra'$ can be used for both continuous and discrete values of $j_3$. We can moreover check that terms corresponding to $D^{j_3,+}$ actually vanish. This is due to 
\bea
\la \prod_{i=1}^3 \Phi^{j_i}_{m_i,\bar{m}_i} \ra'\underset{D^{j_1,+},D^{j_3,+}}{=}0\ , 
\label{ppz}
\eea
which follows from eq. (\ref{sdp}) if we notice that $g^{21}=g^{23}=0$ in this case due to eq. (\ref{ggggp}). This equation holds for generic values of $j_2$, in particular the values $j_2\in-\frac12 +i\R$ which correspond to $C^{j_2}$.  

\paragraph{Case $D^+\times D^+$.} 

The formula (\ref{adsp}) for the product of functions still holds, but the continuous term $\int_{-\frac12+i\R} dj_3\cdots$ vanishes due to eq. (\ref{ppz}). Terms corresponding to $D^{j_3,+}$ representations also vanish by the same argument, but the equation $\la \prod_{i=1}^3 \Phi^{j_i}_{m_i,\bar{m}_i} \ra'\underset{D^{j_1,+},D^{j_2,+}}{=}0$ may fail if a $D^{j_3,-}$ representation is present, due to poles from the factor $K^2$ in eq. (\ref{ppp}). To analyze this matter it is convenient to start with the identity
\bea
\la \prod_{i=1}^3 \Phi^{j_i}_{m_i,\bar{m}_i} \ra'\underset{D^{j_2,+},D^{j_3,-}}{=} -C(j_1,j_2,j_3) K^2 \frac{s(j_{12}^3)^2 s(j_{31}^2)^2 s(j_{23}^1) s(j_{123})}{s(2j_2)s(2j_3)} g^{13}\bar{g}^{13}\ .
\label{pmgg}
\eea
We then send $j_1$ to values corresponding to discrete representations $D^{j_1,+}$. 
Due to momentum conservation we must have $j_1\in j_2-j_3+\Z$. If $j_1\in j_2-j_3-1-\N$ then a double pole from $K^2$ cancels the double zero from $s(j_{13}^2)^2$ and the result is finite. If $j_1\in j_2-j_3+\N$ then the simple pole from $C(j_1,j_2,j_3)$ does not cancel the double zero, and the result vanishes. The formula (\ref{adsp}) therefore reduces to
\bea
\Phi^{j_1}_{m_1,\bar{m}_1}\Phi^{j_2}_{m_2,\bar{m}_2} = \frac{2}{256\pi^2} \sum_{j_3\in j_1+j_2+1+\N} \la \prod_{i=1}^3 \Phi^{j_i}_{m_i,\bar{m}_i} \ra' 2\pi i \Res \frac{1}{R^{j_3}_{m_3,\bar{m}_3}} \Phi^{j_3}_{-m_3,-\bar{m}_3}\ .
\eea

\paragraph{Case $D^+\times D^-$.} 

The formula (\ref{adsp}) for the product of functions still holds, and the analysis of eq. (\ref{pmgg}) in the previous case determines which terms may vanish. Nonvanishing $D^{j_3,+}$ terms occur for $j_3\in j_2-j_1-1-\N\cap (-\frac12,\infty)$ and nonvanishing $D^{j_3,-}$ terms occur for $j_3\in j_1-j_2-1-\N\cap (-\frac12,\infty)$. Depending on the values of $j_1,j_2$ we can have either $D^{j_3,+}$ terms, or $D^{j_3,-}$ terms, or no discrete terms at all in the case $|j_1-j_2|\leq \frac12$.

\zeq\section{Conclusion}

At the level of symmetry algebras, the Wick rotation from $\Hp$ to $AdS_3$ amounts to a map from $s\ell(2,\C)$ to $s\ell(2,\R)\times s\ell(2,\R)$, which can be viewed as two different real forms of the same algebra $s\ell(2,\C)^\C = s\ell(2,\C)\times s\ell(2,\C)$. In particular, the Wick rotation maps the continuous representation $C^j$ of $s\ell(2,\C)$ to the representation $\int_0^1 d\al\ C^{j,\al}\otimes C^{j,\al}$ of $s\ell(2,\R)\times s\ell(2,\R)$. The fact that such an irreducible representation is mapped to a reducible one implies that the symmetry constraints are weaker in $AdS_3$ than in $\Hp$. Namely, the $\Hp$ three-point function should be $\la \prod_{i=1}^3 \Phi^{j_i}_{m_i,\bar{m}_i} \ra = C(j_1,j_2,j_3) H(j_i,m_i,\bar{m}_i)$ where $H$ is determined by $s\ell(2,\C)$ symmetry; while the $AdS_3$ three-point function can in principle be $\la \prod_{i=1}^3 \Phi^{j_i}_{m_i,\bar{m}_i} \ra =C'(j_1,j_2,j_3|\al_1,\al_2,\al_3) H'(j_i,m_i,\bar{m}_i)$ with $m_i,\bar{m}_i\in \al_i+\Z$, where the $s\ell(2,\R)\times s\ell(2,\R)$ symmetry  determines $H'$ but not the $\al_i$-dependence. 

For the full $\Hp$ and $AdS_3$ WZNW models (and not just their minisuperspace limits), the assumption that these models are related by Wick rotation therefore determines part of the $AdS_3$ structure constants (analogs of $C'$) in terms of the $\Hp$ conformal blocks (analogs of $H$). This assumption is therefore rather nontrivial and it should be
carefully justified. The best justification may come a posteriori, if an ansatz for the $AdS_3$ three-point function derived by Wick rotation can be shown to obey crossing symmetry. Such questions did not arise in the minisuperspace limit, as the bases of functions $\Phi^j_{m,\bar{m}}$ on $\Hp$ and $AdS_3$ are related by Wick rotation by definition, and crossing symmetry amounts to the associativity of the product of functions on these spaces. But proving crossing symmetry -- or equivalently the associativity of the operator product expansion -- certainly is the most important and difficult task in solving the $AdS_3$ WZNW model.


\acknowledgments{I am very grateful to Kazuo Hosomichi for decisive help in constructing the $t$-basis of functions on $AdS_3$. Moreover I wish to thank Nicolas Cramp\'e and Sabine Morisset for comments on the draft of this article.}


\begin{thebibliography}{10}

\bibitem{mo00a}
J.~M. Maldacena and H.~Ooguri, {\it Strings in {$AdS_3$ and $SL(2,\mathbb{R})$
  WZW model. I}},  {\em J. Math. Phys.} {\bf 42} (2001) 2929--2960
  [\href{http://arXiv.org/abs/hep-th/0001053}{{\tt hep-th/0001053}}].

\bibitem{tes99}
J.~Teschner, {\it Operator product expansion and factorization in the {$H_3^+$}
  {WZNW} model},  {\em Nucl. Phys.} {\bf B571} (2000) 555--582
  [\href{http://arXiv.org/abs/hep-th/9906215}{{\tt hep-th/9906215}}].

\bibitem{tes01b}
J.~Teschner, {\it Crossing symmetry in the {$H_3^+$} {WZNW} model},  {\em Phys.
  Lett.} {\bf B521} (2001) 127--132
  [\href{http://arXiv.org/abs/hep-th/0108121}{{\tt hep-th/0108121}}].

\bibitem{wit98}
E.~Witten, {\it {Anti-de Sitter space and holography}},  {\em Adv. Theor. Math.
  Phys.} {\bf 2} (1998) 253--291
  [\href{http://arXiv.org/abs/hep-th/9802150}{{\tt hep-th/9802150}}].

\bibitem{tes97b}
J.~Teschner, {\it The mini-superspace limit of the {SL(2,C)/SU(2) WZNW} model},
   {\em Nucl. Phys.} {\bf B546} (1999) 369--389
  [\href{http://arXiv.org/abs/hep-th/9712258}{{\tt hep-th/9712258}}].

\bibitem{hr06}
K.~Hosomichi and S.~Ribault, {\it Solution of the $h_3^+$ model on a disc},
  {\em JHEP} {\bf 01} (2007) 057
  [\href{http://arXiv.org/abs/hep-th/0610117}{{\tt hep-th/0610117}}].

\bibitem{bas07}
D.~Basu, {\it The plancherel formula for the universal covering group of
  sl(2,r) revisited},  \href{http://arXiv.org/abs/arXiv:0710.2224
  [hep-th]}{{\tt arXiv:0710.2224 [hep-th]}}.

\bibitem{kv97}
G.~Kerimov and Y.~Verdiyev, {\it Clebsch-gordan problem for three-dimensional
  lorentz group in the elliptic basis : I. tensor product of continuous
  series},  {\em J. Phys. A: Math. Gen.} {\bf 31} (1998) 3573.

\bibitem{kv98}
G.~Kerimov and Y.~Verdiyev, {\it Clebsch-gordan problem for three-dimensional
  lorentz group in the elliptic basis : Ii. tensor products involving discrete
  series},  {\em J. Phys. A: Math. Gen.} {\bf 32} (1999) 3385.

\bibitem{rib07}
S.~Ribault, {\it Boundary three-point function on ads2 d-branes},  {\em JHEP}
  {\bf 01} (2008) 004 [\href{http://arXiv.org/abs/0708.3028}{{\tt 0708.3028}}].

\bibitem{fh01}
T.~Fukuda and K.~Hosomichi, {\it Three-point functions in sine-liouville
  theory},  {\em JHEP} {\bf 09} (2001) 003
  [\href{http://arXiv.org/abs/hep-th/0105217}{{\tt hep-th/0105217}}].

\end{thebibliography}

\providecommand{\href}[2]{#2}\begingroup\raggedright\endgroup

\end{document}